\documentclass[10pt]{iopart}
\usepackage{iopams}
\usepackage{times}
\usepackage{harvard}

\begin{document}
\title[Quantum gears]{Quantum gears: a simple mechanical system in the
quantum regime}
\author{Angus MacKinnon}
\address{Blackett Laboratory, Imperial College London, Prince Consort Rd, London
SW7 2BW, UK}

\begin{abstract}
The quantum mechanics of a simple mechanical system is considered. A group
of gears can serve as a model for several different systems such as an
artifically constructed nanomechanical device or a
group of ring molecules.  An expression is derived for the quantisation
of the dynamics of a 2--gear system. The general solution for several
gears is discussed.

\end{abstract}



\section{Introduction}
With the rapid rise in interest in micro- and nano-mechanical and electromechanical
devices it will soon be necessary to consider the quantum aspects of the
behaviour of such mechanical devices in the same way as the development
of micro- and nano-scale electronic devices forces us to consider quantum
finite size and interference effects as well as single electron effects
\cite{Drexler92,CR99,Cleland02}.
As a model system to investigate the sort of effects which the onset of
quantum behaviour could give rise to, a system of gears is considered.
Such a system is a vital component of most classical  ``machines'' but one
for which it is easy to anticipate different behaviour in the quantum system.
Classically the angular velocities of the gears are locked together by
the action of the teeth, whereas in the quantum system the angular momenta
of the individual gears will be quantised in a way which is not necessarily
consistent with the expected classical solution.  

The system of 2 gears serves as a model system for any nano--machine in
which circular (or spherical components) are constrained to rotate at angular
frequencies which have a fixed ratio to each other, but are also subject
to quantisation of their angular momenta.  One could consider, for example,
anything involving different sized wheels moving along a surface, such
as a  ``penny--farthing'' bicycle.

For 2 gears with $n_1$ and $n_2$ teeth their angular velocities, $\omega_1$
and $\omega_2$, are locked together as $n_1\omega_1+n_2\omega_2 = 0$.  On
the other hand each of them is subject to an angular momentum quantisation,
$L=I\omega=m\hbar$ where $I$ is the moment of inertia and $m$ is an integer.
By eliminating the 2 $\omega$'s we obtain a constraint on the behaviour
of the pair of gears
\begin{equation}
n_1 m_1 I_1^{-1} = n_2 m_2 I_2^{-1}
\label{eq:locked}
\end{equation}
which has no non--trivial solutions (\mbox{i.e.} with integer $m_1 \& m_2$
other than $m_1=m_2=0$)
except for special cases such as 2 identical gears.  The gears appear to
be locked together such that they cannot rotate.

Is this picture too na{\"\i}ve?  It must be possible to return to the classical
behaviour as the gears become larger; somehow the gears must be unlocked
on macroscopic scales.  
On the other hand there are lots of examples
of systems with quantised angular momenta for which the strict quantisation
is lifted when the systems are combined, such as when 2 atoms form 
a molecule and the previously quantised atomic orbitals are hybridised. 

\section{A Model}
In order to investigate the matter further we consider the simplest Hamiltonian
which describes such a system
\begin{eqnarray}
H&=&{L_1^2\over 2 I_1} + {L_2^2\over 2I_2} + V(n_1\theta_1+n_2\theta_2)\\
&=&{1\over 2 I_1}\left({\hbar\over\rmi}{\partial\over\partial\theta_1}\right)^2
+ {1\over 2 I_2}\left({\hbar\over\rmi}{\partial\over\partial\theta_2}\right)^2
+ V_0 v(n_1\theta_1+n_2\theta_2)
\label{equ:2gears}
\end{eqnarray}
where $\theta_n$ is the angle of the $n$th gear, $0 \le \theta < 2\pi$,
and $v(\theta)$ is a periodic function with period $2\pi$ with a single
minimum with $v=0$ and a single maximum with $v=1$ in each period. $V_0$
is the amplitude of the potential 
which describes the interaction between the gears. When $V_0\rightarrow\infty$
the gears are constrained such that $n_1\theta_1+n_2\theta_2$ is constant,
as required.
It will be useful, however, to consider finite $V_0$ as well.

\Eref{equ:2gears} may be separated by changing to new variables
\begin{eqnarray}
p&=&{I_1\over n_1}\theta_1 - {I_2\over n_2}\theta_2\\
s&=&n_1\theta_1+n_2\theta_2
\end{eqnarray}
which represent the primary and secondary motion of the gears respectively.
Here {\it primary\/} refers to the usual motion expected of gears in which
the angular velocities have a fixed ratio and {\it secondary\/} refers
to the deviation from the primary behaviour.  

In terms of $p$ and $s$ the Hamiltonian may now be rewritten as
\begin{equation}
-{\hbar^2\over 2}\left({n_1^2\over I_1} + {n_2^2\over I_2}\right)
{\partial^2\over\partial p^2}
-{\hbar^2\over 2}\left({I_1\over n_1^2} + {I_2\over n_2^2}\right)
{\partial^2\over\partial s^2}
+ V_0 v(s) ,
\label{equ:new2gears}
\end{equation}

\subsection{Classical behaviour}
The part in $p$ of \eref{equ:new2gears} is the equation of a free system with
an effective moment of inertia, $I_{\mbox{\scriptsize\rm eff}}$, such that 
\begin{equation}
{1\over I_{\mbox{\scriptsize\rm eff}}} = {n_1^2\over I_1} + {n_2^2\over I_2}
.
\end{equation}
The rest of \eref{equ:new2gears} describes a system subject to the potential
$V_0 v(s)$ whose behaviour depends on whether the energy is greater than
or less than $V_0$.  For a classical system in the former case the motion is effectively free,
whereas in the latter case the system oscillates around the minimum of
the potential.  Manufacturers of gears typically try to design the teeth
to minimise this motion.

\subsection{Quantum behaviour}
What changes when we consider the quantum solutions of \eref{equ:new2gears}?
Firstly, the case $V_0=0$: here the solutions are
simple plane waves in both representations and are subject to the boundary
conditions that the wave function is single--valued and continuous at $\theta=0,
2\pi$ giving
\begin{equation}
\exp(\rmi m_1\theta_1)\exp(\rmi m_2\theta_2)
= \exp(k_{\rm p} p)\exp(\rmi k_{\rm s} s) .
\end{equation}
Combining this
with the definitions of $p$ and $s$ and demanding that the equation is valid
for all $(\theta_1,\theta_2)$ gives the quantisation conditions
\begin{eqnarray}
\label{eq:2qnums:p}
k_{\rm p} &= \left( {n_1 n_2\over I_1 I_2} \left(m_1 n_2 - m_2 n_1\right)\right)
\left({n_1^2\over I_1} + {n_2^2\over I_2 }\right)^{-1}\\
k_{\rm s} &= \left( {n_1 m_1\over I_1} + {n_2 m_2 \over I_2}\right)\left( {n_1^2\over I_1} + {n_2^2\over
I_2 } \right)^{-1}
\label{eq:2qnums:s} 
\end{eqnarray}

When $V_0$ is finite the secondary part of \eref{equ:new2gears} is identical
to the Schr{\"o}dinger equation for a 1--dimensional crystal as discussed
in any textbook of solid state physics \cite{AM76}.  The particular results
required here are
\begin{enumerate}
\item\label{BlochsTheorem} the matrix elements of the periodic potential
with different wave--like solutions are non--zero only when the wave vectors
are related by
\begin{equation}
k'_{\rm s}=k_{\rm s}+j .
\end{equation}
In the present case the integral over $p$ in the evaluation of the matrix
elements is non--zero only for terms diagonal in $k_{\rm p}$.
\item it follows that $k_{\rm s}$ values in the range $-{\textstyle\frac12} < k_{\rm s}
\le +{\textstyle\frac12}$ are good quantum numbers representing independent
solutions (\mbox{i.e.} they are in the 1st Brillouin zone). $k_{\rm p}$ is always
a good quantum number.
\item\label{gvel} the group velocity $\rmd E/\rmd k_{\rm s}$, where $E$ is the energy, is identically zero
for $k_{\rm s} = m{\textstyle\frac12}$ where $m$ is an integer (\mbox{i.e.} at the
middle and edges of the Brillouin zone).
\item\label{flatbands} in the limit $V_0\gg E$ the states in each unit cell are decoupled
from one another and the group velocity tends to zero.
\end{enumerate}

Using result (\ref{BlochsTheorem}) and \eref{eq:2qnums:s} we can write
\begin{equation}
{n_1 m'_1\over I_1} + {n_2 m'_2 \over I_2} 
= {n_1 m_1 \over I_1} + {n_2 m_2 \over I_2} 
+ j\left({n_1^2\over I_1} + {n_2^2 \over I_2}\right)
\label{eq:nonvector}
\end{equation}
It is useful to visualise this by considering vectors on a 2D rectangular
lattice with
basis vectors $(I_1^{-1/2}, I_2^{-1/2})$ so that \eref{eq:nonvector} may
be expressed as
\begin{equation}
\bi{m}'\cdot\bi{n} = \bi{m}\cdot\bi{n} + j\bi{n}^2
\label{eq:projections}
\end{equation}
where the vectors $\bi{n}$ represent points on the lattice with the integer
components $(n_1,n_2)$ and the terms in \eref{eq:projections} represent
projections of the vectors onto $\bi{n}$.  The simple solution of \eref{eq:projections}
for $\bi{m}'$ is given by 
\begin{equation}
\label{eq:trivialsolution}
\bi{m}' = \bi{m} + j\bi{n}
\end{equation}
but other solutions may be found by adding a lattice vector perpendicular
to $\bi{n}$.  Note, however, that $k_{\rm p}$ contains $\bi{m}\times\bi{n}$ so
that the addition of such a vector would change the value of $k_{\rm p}$.  Thus
\eref{eq:trivialsolution} is in fact the complete solution.

The general solution for the wave function of the two gear system 
can therefore be written as
\begin{equation}
\psi_{\bi{m}}(p,s) = \exp\left[\rmi k_{\rm p}(\bi{m}) p\right]
\sum_j a_j \exp\left[\rmi k_{\rm s}(\bi{m}+j\bi{n}) s\right]
\end{equation}
where the solutions are distinct for all $\bi{m}$ not related by \eref{eq:trivialsolution}.

For hard gears ($V_0\rightarrow\infty$) using (\ref{flatbands}) above the
group velocity of the secondary part is zero; the teeth of the gears do
not tunnel through each other.  Apart from the
zero point energy of the secondary part the energy of the system may be
written
\begin{equation}
E = -{\hbar^2\over 2}\left(n_1 n_2\over I_1 I_2\right)^2\left({n_1^2\over I_1} + {n_2^2\over I_2}\right)^{-1}
(m_1 n_2 - m_2 n_1)^2
\label{eq:2qnums:E}
\end{equation}

\subsection{Soft gears}
When $V_0$ is finite it becomes possible for the teeth of the gears to
tunnel through one another.  One might visualise this in terms of the interaction
between 2 ring molecules where the effective value of $V_0$ would depend
on their separation.

For such gears to be useful the group velocity of the secondary part should
be zero, which from (\ref{gvel}) above 
can only be guaranteed for $k_{\rm s}$ at the centre and edges of the Brillouin
zone, where $k_{\rm s}$ is an integer multiple of $\frac12$ or 
\begin{equation}
\bi{m}\cdot\bi{n} = \frac{j}2\bi{n}^2 .
\end{equation}
When $I_1=I_2$ and $n_1=n_2$ there
are many such solutions, but when $(I_1/I_2)^{1/2}$ is irrational there are none
at all.

\section{Many Gears --- A Quantum Machine}
We define a quantum machine as a collection of gears all interlocking with
each other, either directly or indirectly, such that, classically, if one
is rotating then all are.  There are cases, such as 3 mutually interlocking
gears, which are classically frustrated but which may have quantum solutions
involving tunnelling of the teeth. Such systems will not be considered here.

An $N$-gear system will have a wave function which may be written in the
form $\psi(p,s_1,\ldots,s_{N-1})$ where
\begin{eqnarray}
p   &=& \sum_{i=1}^N a_i \theta_i\\
s_i &=& n_i \theta_i + n_{i+1} \theta_{i+1}\;.
\label{eq:Ncoords}
\end{eqnarray}
The 2nd derivatives in the kinetic energy terms of the Hamiltonian may then be written as
\begin{equation}
\begin{array}{ccc}
\displaystyle{\partial^2\over\partial\theta_1^2}
&=& \left(a_1 \displaystyle{\partial\over\partial p} + n_1 \displaystyle{\partial\over\partial s_1}\right)^2\\
\vdots&&\vdots\\
\displaystyle{\partial^2\over\partial\theta_i^2}
&=& \left(a_i \displaystyle{\partial\over\partial p} + n_i \displaystyle{\partial\over\partial s_{i-1}}
+ n_i \displaystyle{\partial\over\partial s_i}\right)^2\\
\vdots&&\vdots\\
\displaystyle{\partial^2\over\partial\theta_N^2}
&=& \left(a_N \displaystyle{\partial\over\partial p} + n_N \displaystyle{\partial\over\partial s_{N-1}}\right)^2
\end{array}
\label{eq:NKEs}
\end{equation}
For the $p$ part of the equation to be separable the coefficients of the
partial derivatives of the form $\partial^2/\partial p\partial s_i$ should
be zero. This requires that
\begin{equation}
{a_i n_i\over I_i} + {a_{i+1} n_{i+1}\over I_{i+1}} = 0
\end{equation}
which has a solution 
\begin{equation}
a_i = (-1)^i {I_i\over n_i}\;.
\end{equation}
Let us now assume that the general solution may be written in the form
\begin{equation}\fl
\exp(\rmi k_{\rm p} p)f(s_1,\ldots,s_{N-1}) 
= \sum_{m_1,\ldots,m_N} c_{m_1,\ldots,m_N}\prod_{i=1}^N \exp(\rmi m_i\theta_i)
\end{equation}
such that the primary part may be separated and takes a simple wave form
and the total function may be written as a sum over every possible rotational
state of the gears.

The coefficient $c$ for a particular set of $m$'s may be calculated using
\begin{equation}\fl
(2\pi)^N c_{m_1,\ldots,m_N} 
= \prod_i^N\int_0^{2\pi}\rmd\theta_i \exp(-\rmi m_i\theta_i) \exp(\rmi k_{\rm p} p)f(s_1,\ldots,s_{N-1})\;.
\label{eq:int-theta}
\end{equation}
To evaluate this integral it is useful to transform it into the $p$ and
$s_i$ representation by writing
\begin{eqnarray}
\fl
\left[\begin{array}{c} p\\ s_1 \\ \vdots\\ s_{N-1}\end{array}\right]
&=&\left[\begin{array}{cccc} -(I_1/n_1) & (I_2/n_2) & \cdots & (-1)^N (I_N/n_N)\\
n_1 & n_2 &&\\
&\ddots&\ddots&\\
&&n_{N-1}&n_N
\end{array}\right]
\left[\begin{array}{c} \theta_1\\ \theta_2 \\ \vdots \\ \theta_N \end{array}\right]
\label{eq:ps-theta}\\
\bi{p}&=&{\mathbf T}\btheta\nonumber
\end{eqnarray}
so that the \eref{eq:int-theta} becomes
\begin{equation}\fl
c_{\bi{m}} \propto \int_{-\infty}^{+\infty} \rmd p 
\prod_{i=1}^{N-1}\int\rmd s_i
\exp\left(-\rmi\:\bi{m}\cdot{\mathbf T}^{-1}\bi{p}\right)
\exp(\rmi k_{\rm p} p)f(s_1,\ldots,s_{N-1})
\label{eq:int-ps}
\end{equation}
where $\bi{m}$ and $\bi{p}$ are vectors with components $(m_1,\ldots,m_N)$
and $(p,s_1,\ldots)$ respectively and we can extend the limits on the integral
to $\pm\infty$ without loss of generality, as long as it is understood that
the the intergral is
normalized by division by its range.

The integral over $p$ in \eref{eq:int-ps} is zero unless
\begin{equation}
k_p = \sum_i^N m_i \left[{\mathbf T}^{-1}\right]_{i1}\;.
\label{eq:kp}
\end{equation}
In order to evaluate ${\mathbf T}^{-1}$ consider the equation $\sum_j T_{ij}\tau_j
= \delta_{i1}$.  All but the 1st row may be rewritten
\begin{equation}
\tau_i = -\frac{n_{i-1}}{n_i}\tau_{i-1} = (-1)^{i-1}\frac{n_1}{n_i}\tau_1 \;.
\end{equation}  
Substituting this into the 1st row gives
\begin{equation}
\sum_{i=1}^N (-1)^i \frac{I_i}{n_i} \tau_i = -\sum_{i=1}^N \frac{I_i}{n_i^2}
n_1\tau_1 = 1
\end{equation}
which leads to the result
\begin{equation}
\tau_i = \frac{(-1)^i}{n_i}/\sum_{j=1}^N \frac{I_j}{n_j^2}
\end{equation}
and from \eref{eq:kp}
\begin{equation}
k_p = \left(\sum_{i=1}^N (-1)^i \frac{m_i}{n_i}\right)/\left(\sum_{i=1}^N \frac{I_i}{n_i^2}\right)
\label{eq:kp2}
\end{equation}
Using \eref{eq:NKEs} the kinetic energy of the primary motion of the system
may be evaluated as
\begin{eqnarray}
E&=& -\frac{\hbar^2}2 \left(\sum_{i=1}^N \frac{I_i}{n_i^2}\right) 
{\partial^2\over\partial p^2}\\
&=& \frac{\hbar^2}2 \left(\sum_{i=1}^N (-1)^i \frac{m_i}{n_i}\right)^2/\left(\sum_{i=1}^N \frac{I_i}{n_i^2}\right)\;.
\label{eq:pKE}
\end{eqnarray}
Compare these results with \eref{eq:2qnums:p} and \eref{eq:2qnums:E}.

\section{Conclusions}
The quantum  mechanics of a system of interlocking gears has been considered
as a model for a nano--machine.  It has been shown that the locking of
the gears tentatively predicted in the introduction does not occur and
an expression has been derived for the quantised energy of 2 gears. The
case of $N$ interlocking gears has been considered and a general exporession
for the quantisation of the primary motion has been presented.

When the potential describing the interaction of the teeth is softened
most solutions involve tunnelling of the teeth.

\ack The author would like to acknowledge useful discussions with Andrew
Armour and the hospitality of the Cavendish Laboratory, Cambridge where
this work was carried out.

\section*{References}

\end{document}